\documentclass[journal,onecolumn,12pt]{IEEEtran}
\usepackage{pslatex}
\usepackage{amsfonts,color,morefloats}
\usepackage{amssymb,amsmath,latexsym,amsthm}
\usepackage{hyperref}

\newcommand{\wt}{{\mathrm{wt}}}

\newcommand{\tr}{{\mathrm{Tr}}}
\newcommand{\gf}{{\mathrm{GF}}}

\newcommand{\C}{{\mathcal{C}}}

\newtheorem{theorem}{Theorem}
\newtheorem{lemma}[theorem]{Lemma}

\newtheorem{corollary}[theorem]{Corollary}

\newtheorem{example}{Example}

\begin{document}

\title{Binary linear codes with at most $4$ weights\thanks{The research is
supported by  a National Key Basic Research Project of China (2011CB302400), National Science Foundation of China (61379139), the ``Strategic Priority Research Program" of the Chinese Academy of Sciences, Grant No. XDA06010701 and  Foundation of NSSFC(No.13CTJ006).}}

\author{Fei Li\thanks{F. Li is with the School of Statistics and Applied Mathematics,   Anhui University of Finance and Economics,
 Bengbu City,  Anhui Province, {\rm 233030}, China Email:cczxlf@163.com},
Yang Yan$^{*}$\thanks{ Y. Yan is the corresponding author and he is with the National Engineering Laboratory for Information Security Technologies
,  Institute of Information Engineering,
             Chinese Academy of Sciences,
              Beijing, {\rm 100195},  China Email:yanyang9021@iie.ac.cn},
              Qiuyan Wang \thanks{Q. Wang is with the  State Key Laboratory of Information Security, Institute of Information Engineering, Chinese Academy of Sciences, Beijing 100195,  China Email:wangyan198801@163.com} and
              Tongjiang Yan \thanks{T. Yan is with the College of Science, China University of Petroleum, Qingdao 266580, China Email:yantoji@163.com }
}

\date{\today}
\maketitle

\begin{abstract}
For the past decades, linear codes with few weights have been widely studied, since they have applications in space communications, data storage and cryptography. In this paper, a class of binary linear codes is constructed and their weight distribution is determined. Results show that they are at most 4-weight linear codes. Additionally, these codes can be used in secret sharing schemes.

\end{abstract}


\section{Introduction and main results}\label{sec-intro}

Throughout this paper, let $q=2^{m}$ for a positive integer $m$.
Denote $ \mathbb{F}_q =\mathbb{F}_{2^{m}} $ the finite field with $ q $
elements and $\mathbb{F}_q^{*}$ the multiplicative group of $\mathbb{F}_q$.

Let $\mathbb{F}_2^{n}$ denote the vector space of all $n$-tuples over the binary field $\mathbb{F}_2$. A binary code $\C$ of length $n$ is a subset of $\mathbb{F}_2^{n}$. Usually, the vectors in $\C$ are called codewords of $\C$. For codewords $\mathbf{x}$ and $\mathbf{y}\in \C$, the distance $d(\mathbf{x},\mathbf{y})$ is referred as the number of coordinates in which $\mathbf{x}$ and $\mathbf{y}$ differ. The (Hamming) distance of a code $\C$ is the smallest distance between distinct codewords and is an important invariant. An $[n,k,d]$ binary linear code $\C$ is defined as a $k$-dimensional subspace of $\mathbb{F}_2^{n}$ with distance $d$.

For a codeword $\mathbf{c}\in \C$, the (Hamming) weight $wt(\mathbf{c})$ is the number of nonzero coordinate in $\mathbf{c}$. We use $A_{i}$ to denote the number of codewords of weight $i$ in $\C$. Then $(1, A_1,\cdots,A_{n})$ is called the weight distribution of $\C$. And the weight enumerator is defined to be the polynomial $1+A_1x+A_2x^{2}+\cdots+A_{n}x^{n}$.
If the number of nonzero $A_{i}$ ($1\leq i\leq n$)  equals $ t$,  then $\C$ is called a $t$-weight code.
Readers can refer to \cite{HP03} for a general theory of linear codes.

The weight distribution is an important research topic in coding theory, as it contains crucial information to compute the probability of error correcting and detection.  A great deal of researchers are devoted to construct and determine specific linear codes \cite{CKNC12,DLLZ15,LYL14,WDX15}. The weight distribution of Reed--Solomon codes were determined by Blake \cite{BK91} and   Kith \cite{K89}.  A survey of the hamming weights in irreducible cyclic codes was given by Ding and Yang in \cite{DY13}.  The weight distributions of reducible cyclic codes could be found in \cite{DLMZ11,F12,FL08,FLX12,LF08,YCD06}. Recently, Ding \cite{Ding15,DLN08} proposed a generic construction of linear codes as follows.

Let $D=\{d_1,d_2,\ldots, d_n\} \subseteq \mathbb{F}_{q}^{*}$ and $\tr$ denote the
trace function from $\mathbb{F}_q$ to $\mathbb{F}_2. $
Linear codes $\C_D$ of length $n$ can be constructed by
$$
\C_{D}=\{\left(\tr(xd_1), \tr(xd_2),\ldots, \tr(xd_{n})\right):x\in \mathbb{F}_{q}\}.
$$
Here  $D$ is called the defining set of $ \C_{D}$. The dimension of the code $\C_D$ have been presented in \cite{DH07} and is equal to the dimension of the $\mathbb{F}_2$-linear space of $\mathbb{F}_{q}$ spanned by $D$.   This method has been  widely  used by some researchers to acquire linear codes with few weights \cite{Ding09,DD14,DD15,DGZ13,WDLX15,ZD14,ZLFH15}. In this paper, we will present a class of binary linear codes with at most four weights.

For $a\in \mathbb{F}_{2}$, we  set
\begin{equation}\label{eq-D}
D_{a}=\{x\in \mathbb{F}_q^{*}: \tr(x)=a\}.
 \end{equation}
 Motivated by the research work in \cite{DD14}, a class of   binary linear codes $\C_{D_a}$ is defined by
\begin{eqnarray}\label{defcode}
         \C_{D_a}=\left\{\left(\tr(xd^{2^{h}+1})\right)_{d\in D_a}: x\in \mathbb{F}_q \right\},
\end{eqnarray}
where $a\in \mathbb{F}_2$ and  $h<m$ is a   positive factor of $m$. The weight distribution of the presented linear codes is settled and the main
results are listed as follows.

\begin{table}[ht]
\centering
\caption{The weight distribution of the codes of Theorem \ref{thm1} }\label{tab1}
\begin{tabular}{|l|l|}
\hline
\textrm{Weight} $w$ \qquad& \textrm{Multiplicity} $A$   \\
\hline
0 \qquad&   1  \\
\hline
$2^{m-2}$ \qquad&  $2^{m}-1-2^{m-h}$  \\
\hline
$2^{m-2}-2^{\frac{m+h-4}{2}}$  \qquad& $2^{m-h-1}+2^{\frac{m-h-2}{2}}$  \\
\hline
$2^{m-2}+2^{\frac{m+h-4}{2}}$ \qquad&    $2^{m-h-1}-2^{\frac{m-h-2}{2}}$ \\
\hline
\end{tabular}
\end{table}

\begin{theorem}\label{thm1}
Let $m/h$ be odd. Then the code $\C_{D_{0}}$ defined in \eqref{defcode} is a $[2^{m-1}-1, m]$ binary
linear code with weight distribution in \autoref{tab1}.
\end{theorem}

\begin{theorem}\label{thm2}
Let $m/h$ be odd. Then the code $\C_{D_{1}}$
defined in \eqref{defcode} is a $[2^{m-1}, m]$ binary linear code with weight distribution in \autoref{tab2}.
\end{theorem}

\begin{table}[h!]
\centering
\caption{The weight distribution of the codes of Theorem \ref{thm2}}\label{tab2}
\begin{tabular}{|l|l|}
\hline
\textrm{Weight} $w$ \qquad& \textrm{Multiplicity} $A$   \\
\hline
0 \qquad&   1  \\
\hline
$2^{m-2}$ \qquad&  $2^{m}-1-2^{m-h}$  \\
\hline
$2^{m-2}-2^{\frac{m+h-4}{2}}$  \qquad& $2^{m-h-1}-2^{\frac{m-h-4}{2}}$  \\
\hline
$2^{m-2}+2^{\frac{m+h-4}{2}}$ \qquad&    $2^{m-h-1}+2^{\frac{m-h-4}{2}}$ \\
\hline
\end{tabular}
\end{table}

\begin{table}[ht]
\centering
\caption{The weight distribution of the codes of Theorem \ref{thm3}.}\label{tab3}
\begin{tabular}{|c|c|}
\hline
\textrm{Weight} $w$ \qquad& \textrm{Multiplicity} $A$   \\
\hline
0 \qquad&   1  \\
\hline
$2^{m-2}+(-1)^{\frac{e}{h}}2^{e+h-1}$ \qquad&  $\frac{2^{m-2h-1}-1-(-1)^{\frac{e}{h}}2^{e-h-1}}{2^{h}+1}$  \\
\hline
$2^{m-2}+(-1)^{\frac{e}{h}}2^{e+h-2}$  \qquad& $(2^{h}-1)2^{m-2h}$  \\
\hline
$2^{m-2}$ \qquad&    $2^{m-1}-(-1)^{\frac{e}{h}}(2^{h}-1)(2^{m-2h-1}+2^{e-h-1})$ \\
\hline
$2^{m-2}-(-1)^{\frac{e}{h}}2^{e-1}$ \qquad&    $\frac{2^{m+h-1}+2^{m-2h-1}-2^{m-1}-2^{h}+(-1)^{\frac{e}{h}}(2^{e+h-1}+2^{m-1}-2^{m-2h-1})}{2^{h}+1}$ \\
\hline
\end{tabular}
\end{table}

The above two theorems present the parameters of $\C_{D_a}$ $(a=0,1)$ of \eqref{defcode} for the case that $m/h\equiv1\pmod{2}$. Next, we will assume $m/h$ is even and $m=2e$.  In this case, the parameters of $\C_{D_a}$ $(a=0,1)$  of \eqref{defcode} are given in the following two theorems.
\begin{theorem}\label{thm3}
Let $m/h$ be even and $m/h>2$. Then the code $\C_{D_{0}}$
defined in \eqref{defcode} is a $[2^{m-1}-1, m]$ binary linear code with weight distribution in \autoref{tab3}.
\end{theorem}

\begin{table}[ht]
\centering
\caption{The weight distribution of the codes of Theorem \ref{thm4}}\label{tab4}
\begin{tabular}{|l|l|}
\hline
\textrm{Weight} $w$ \qquad& \textrm{Multiplicity} $A$   \\
\hline
0 \qquad&   1  \\
\hline
$2^{m-2}+(-1)^{\frac{e}{h}}2^{e+h-1}$ \qquad&  $\frac{2^{m-2h-1}+(-1)^{\frac{e}{h}}2^{e-h-1}}{2^{h}+1}$  \\
\hline
$2^{m-2}+(-1)^{\frac{e}{h}}2^{e+h-2}$  \qquad& $(2^{h}-1)2^{m-2h}$  \\
\hline
$2^{m-2}$ \qquad&    $2^{m-1}-1+(-1)^{\frac{e}{h}}2^{e-h-1}(2^{h}-1)-(2^{h}-1)2^{m-2h-1}$ \\
\hline
$2^{m-2}-(-1)^{\frac{e}{h}}2^{e-1}$ \qquad&    $\frac{(2^{e}-(-1)^{\frac{e}{h}})2^{e+h-1}}{2^{h}+1}$ \\
\hline
\end{tabular}
\end{table}

\begin{theorem}\label{thm4}
Let $m/h$ be even and $m/h>2$. Then the code $\C_{D_{1}}$
defined in \eqref{defcode} is a $[2^{m-1}, m]$ binary linear code with weight distribution in \autoref{tab4}.
\end{theorem}
Let $ D=\mathbb{F}_{q}^{*}$. If $m/h$ is odd, then $\gcd(2^{h}+1,2^{m}-1)=1$ (Lemma 2.1, \cite{C99}) and it is straightforward to verify that $\C_D $ of \eqref{defcode} is a constant binary  linear code.
If $m/h$ is even and $m>2$, the code  $\C_D$ of \eqref{defcode} is a $2$-weight binary linear code, and the weight distribution of $\C_{D}$ is given in Theorem \ref{thm5}.

\begin{theorem}\label{thm5}
Let $m/h$ be even, $m>2$ and $ D=\mathbb{F}_{q}^{*} $. Then the code $\C_{D}$
defined in \eqref{defcode} is a $[2^{m}-1, m]$
binary linear code with weight distribution in \autoref{tab5}.
\end{theorem}
\begin{table}[ht]
\centering
\caption{The weight distribution of the codes of Theorem \ref{thm5}.}\label{tab5}
\begin{tabular}{|l|l|}
\hline
\textrm{Weight} $w$ \qquad& \textrm{Multiplicity} $A$   \\
\hline
0 \qquad&   1  \\
\hline
$2^{m-1}-(-1)^{\frac{e}{h}}2^{e-1}$ \qquad&  $\frac{(2^{m}-1)2^{h}}{2^{h}+1}$  \\
\hline
$2^{m-1}+(-1)^{\frac{e}{h}}2^{e+h-1}$  \qquad& $\frac{2^{m}-1}{2^{h}+1}$  \\
\hline
\end{tabular}
\end{table}

If $m/h$ is even,  by Lemma 2.1 in \cite{C99}, we know $\gcd(2^{h}+1,2^{m}-1)=2^{h}+1$, i.e., $2^{h}+1\mid 2^{m}-1$. Hence $f(x)=x^{2^{h}+1}$ is a $(2^{h}+1)-$to$-1 $ function over $\mathbb{F}_{q}^{*}$ in the case that $m/h\equiv0\pmod{2}$. This implies that a binary code  may be punctured from the code $\C_{D}$ in Theorem \ref{thm5}.

Let $\overline{D}=\{x^{2^{h}+1}:x\in \mathbb{F}_{q}^{*}\}$ and
\begin{equation}\label{defcode2}
{\C}_{\overline{D}}=\left\{\left(\tr(xd)\right)_{d\in \overline{D}}:x\in\mathbb{F}_{q}\right\}.
\end{equation}
Then the parameters of $\C_{\overline{D}}$ of \eqref{defcode2} can be easily derived from the code $\C_{D}$ in Theorem \ref{thm5}, and  are given in the following corollary.

\begin{corollary}\label{corollary6}
Let $m/h$ be even and $m>2$. Then the code $\C_{\overline{D}}$
defined in \eqref{defcode2}  is a $[\frac{2^{m}-1}{2^{h}+1}, m]$
binary linear code with weight distribution in \autoref{tab6}.
\end{corollary}
\begin{table}[ht]
\centering
\caption{The weight distribution of the codes of Corollary \ref{corollary6}}\label{tab6}
\begin{tabular}{|l|l|}
\hline
\textrm{Weight} $w$ \qquad& \textrm{Multiplicity} $A$   \\
\hline
0 \qquad&   1  \\
\hline
$\frac{2^{m-1}-(-1)^{\frac{e}{h}}2^{e-1}}{2^{h}+1}$ \qquad&  $\frac{(2^{m}-1)2^{h}}{2^{h}+1}$  \\
\hline
$\frac{2^{m-1}+(-1)^{\frac{e}{h}}2^{e+h-1}}{2^{h}+1}$  \qquad& $\frac{2^{m}-1}{2^{h}+1}$  \\
\hline
\end{tabular}
\end{table}
\begin{example}
Let $(m,h)=(5,1)$. For $a=0$, the code $\C_{D_0}$ in Theorem \ref{thm1} has parameters $[15,5,6]$ with weight distribution enumerator $1+10x^{6}+15x^{8}+6x^{10}$. For $a=1$, the code $\C_{D_1}$ in Theorem \ref{thm2} has parameters $[16,5,8]$ with weight enumerator $1+6x^{6}+15x^{8}+10x^{10}$.
\end{example}
\begin{example}
Let $(m,h)=(8,2)$. For $a=0$, the code $\C_{D_0}$ in Theorem \ref{thm3} has parameters $[127,8,56]$ with weight distribution enumerator $1+108x^{56}+98x^{64}+48x^{80}+x^{96}$. For $a=1$, the code $\C_{D_1}$ in Theorem \ref{thm4} has parameters $[128,8,56]$ with weight enumerator $1+96x^{56}+109x^{64}+48x^{80}+2x^{96}$.
\end{example}
\begin{example}
Let $(m,h)=(6,1)$. Then the code $\C_{D}$ in Theorem \ref{thm5} has parameters $[63,6,24]$ with weight enumerator
$1+21x^{24}+42x^{36}$. The code $\C_{\overline{D}}$ in Corollary \ref{corollary6} has parameters $[21,6,8]$ with weight enumerators $1+21x^{8}+42x^{12}$.
\end{example}

\section{Preliminaries}

In this section, we  present some results on Weil sums, which will be needed in calculating the weight distribution of the codes defined in \eqref{defcode}.

An additive character of $\mathbb{F}_{q}$ is a group homomorphism $\chi$ from $\mathbb{F}_{q}$ to
unit circle of the complex plane. Each additive character can be defined as a mapping
$$
\chi_{b}(c)=(-1)^{\tr(bc)} \ \textrm{for\ all }\ c\in \mathbb{F}_{q},
$$
with some $b\in \mathbb{F}_{q}.$ For $b=0$, the additive character
$\chi_0$ is called \textit{trivial}  and the other
characters $\chi_{b}$ with $b\in \mathbb{F}_{q}^{*}$ are called \textit{nontrivial}. For $b=1$,  the character $\chi_{1}$ is called the \textit{canonical additive character} of $\mathbb{F}_{q}$. And it is well-known that $\chi_{b}(x)=\chi_{1}(bx)$ for all $x\in \mathbb{F}_{q}$ \cite{LN97}.

 Define the Weil sum
$$S_{h}(a,b)=\sum_{x\in \mathbb{F}_{q}}\chi_1\left(ax^{2^{h}+1}+bx\right)$$
 where $ a\in\mathbb{F}_{q}^{*}$ and $b \in\mathbb{F}_{q}$. In this paper, we restrict  that $h$ is a proper  positive divisor of $m$. Generally, to evaluate an exponential sum over a finite field is a challenge task. At present, it has been determined only in certain cases \cite{C79,C80,C98,C99,FL08,H07}. Among them is the following cases of $S_h(a,b)$.

\begin{lemma}[\cite{C99}, Theorem 4.1]\label{lem5}
If  $m/h$ is odd, then $\sum_{x\in \mathbb{F}_{q}}\chi_1\left(ax^{2^{h}+1}\right)=0 $
for each $a\in\mathbb{F}_{q}^{*}. $
\end{lemma}

\begin{lemma}[\cite{C99}, Theorem 4.2,]\label{lem6}
Let $b\in\mathbb{F}_{q}^{*} $ and suppose $m/h$ is odd. Then
$S_{h}(a,b)=S_{h}(1,bc^{-1}), $ where $c\in\mathbb{F}_{q}^{*}$ is the unique element satisfying
$ c^{2^{h}+1}=a. $ Further we have
$$
S_{h}(1,b)=\left\{\begin{array}{ll}
                                                                          0, & \textrm{if\ } \ \tr_{h}(b)\neq 1, \\
                                                                          \pm 2^{\frac{m+h}{2}}, & \textrm{if\ } \ \tr_{h}(b)= 1,
                                                                        \end{array}
                                                                        \right.
$$
where and hereafter $\tr_{h}$ is the trace function from $\mathbb{F}_{q}$ to $\mathbb{F}_{2^{h}}.$
\end{lemma}
\begin{lemma} [\cite{C99}, Theorem 5.2]\label{lem7}
Let $m/h$ be even and $m=2e$ for some integer $e$.  Then
$$
S_{h}(a,0)=\left\{\begin{array}{ll}
(-1)^{\frac{e}{h}}2^{e}, & \textrm{if\ } \ a\neq g^{t(2^{h}+1)}  \textrm{\ for any integer\ } t, \\
-(-1)^{\frac{e}{h}}2^{e+h}, & \textrm{if\ } \ a= g^{t(2^{h}+1)}  \textrm{\ for some integer\ } t,
\end{array}
\right.
$$
where $g$ is a generator of $\mathbb{F}_{q}^{*}. $
\end{lemma}

\begin{lemma}[\cite{C99}, Theorem 5.3]\label{lem8}
Let $b\in\mathbb{F}_{q}^{*} $ and suppose $m/h$ is even so that $m=2e $
for some integer $ e.$ Let $ f(x)=a^{2^{h}}x^{2^{2h}}+ax \in \mathbb{F}_{q}[x]. $ There are two cases.
\begin{enumerate}

\item If $a\neq g^{t(2^{h}+1)}$ for any integer  $t$ then $f$ is a permutation polynomial of
$\mathbb{F}_{q}. $ Let $x_{0}$ be the unique element satisfying $f(x)=b^{2^{^{h}}}.$ Then
$$S_{h}(a,b)
=(-1)^{\frac{e}{h}}2^{e}\chi_1\left(ax_{0}^{2^{h}+1}\right).$$
\item If $a= g^{t(2^{h}+1)}$ then $S_{h}(a,b)=0$ unless the equation
$f(x)=b^{2^{^{h}}} $ is solvable. If the equation is solvable, with solution $ x_{0}$ say, then
$$S_{h}(a,b)
=\left\{\begin{array}{ll}
                                                                           -(-1)^{\frac{e}{h}}2^{e+h}\chi_1\left(ax_{0}^{2^{h}+1}\right), & \textrm{if\ } \ \tr_{h}(a)= 0, \\
                                                                           (-1)^{\frac{e}{h}}2^{e}\chi_1\left(ax_{0}^{2^{h}+1}\right),  & \textrm{if\ } \ \tr_{h}(a)\neq 0,
                                                                         \end{array}
                                                                         \right.$$
\end{enumerate}
where $\tr_{h}$ is the trace function from $\mathbb{F}_{q}$ to $\mathbb{F}_{2^{h}}.$
\end{lemma}
\begin{lemma}[\cite{C99}, Theorem 3.1]\label{lem3.1}
Let $g$ be a primitive element of $\mathbb{F}_{q}$. For any $a\in\mathbb{F}_{q}^{*}$ consider the equation $a^{2^{h}}x^{2^{2h}}+ax=0$ over $\mathbb{F}_{q}$.
\begin{enumerate}
\item If $m/h$ is odd then there are $2^{h}$ solutions to this equation for any choice of $a\in\mathbb{F}_{q}^{*}$.
\item If $m/h$ is even then there are two possible cases. If $a=g^{t(2^{h}+1)}$ for some $t$, then there are $2^{2h}$ solutions to the equation. If $a\neq g^{t(2^{h}+1)}$ for any $t$ then there exists one solution only, $x=0$.
\end{enumerate}

\end{lemma}

\section{The proofs of the main results}\label{sec-proof}

We follow the notations fixed in Sect. 2. In this section, we will  determine the length of the code $\C_{D_a}$ $(a=0,1)$ of \eqref{defcode}, and give a formula on the weight of a codeword $\mathbf{c}_{b}$ $(b\in\mathbb{F}_{q}^{*})$ in $\C_{D_a}$ $(a=0,1)$ of \eqref{defcode}. Then we give the proofs of Theorems \ref{thm1} and \ref{thm3}.

By the definition of $D_a$ $(a=0,1)$ in \eqref{eq-D}, we know
$$
|D_{a}|=\left\{\begin{array}{ll}
2^{m-1}-1, & \textrm{if\ } \ a=0, \\
2^{m-1}, & \textrm{if\ } \ a=1.
\end{array}
\right.
$$

Define $N(a,b)=\{x\in \mathbb{F}_{q}: \tr(x)=a\textrm{ and } \tr(bx^{2^{h}+1})=0\}.$
We use  $ wt(c_b)$ to denote the Hamming weight of the  codeword
 $
 \mathbf{c}_b
 $ with $b\in \mathbb{F}_{q}^{*}$
 of the code $\C_{D_{a}}$ $(a=0,1)$ defined in \eqref{defcode}.  It can be easily  checked  that
\begin{equation} \label{eq-wt}
wt(\mathbf{c}_{b})=2^{m-1}-|N(a,b)|.
\end{equation}

In terms of exponential sums, for $b\in \mathbb{F}_{q}^{*},$ we have
\begin{align}\label{eq-weight}
|N(a,b)| &= 2^{-2}\sum_{x\in \mathbb{F}_{q}}\left(\sum_{y \in F_{2}}(-1)^{y\tr(x)-ya}\right)
\left(\sum_{z \in F_{2}}(-1)^{z\tr(bx^{2^{h}+1})}\right)  \nonumber \\
 &=2^{-2}\sum_{x\in \mathbb{F}_{q}}\left(1+(-1)^{\tr(x)-a}\right)
\left(1+(-1)^{\tr(bx^{2^{h}+1})}\right)\nonumber \\
&= 2^{m-2}
+ 2^{-2}\sum_{x\in \mathbb{F}_{q}}(-1)^{\tr(bx^{2^{h}+1})} \nonumber
\quad +
2^{-2}\sum_{x\in \mathbb{F}_{q}}(-1)^{\tr(x+bx^{2^{h}+1})-a} \nonumber\\
&= 2^{m-2} + 2^{-2}\left(S_{h}(b,0)+(-1)^{a}S_{h}(b,1)\right). \nonumber\\
\end{align}

Based on the discussion above, the weight distribution of $\C_{D_a}$ of \eqref{defcode} can be determined by the value distribution of $S_h(b,c)$ with $b\in \mathbb{F}_{q}^{*}$ and $c\in \mathbb{F}_{2}$. Combining \eqref{eq-weight} and the lemmas in preliminaries, we are ready to compute the weight distribution of the codes $\C_{D_a}$ $(a=0,1)$ defined in \eqref{defcode}.\\
\vskip 2mm
\noindent{\textbf{Proof of Theorem \ref{thm1}.}}
By Lemma \ref{lem5}, we have $ S_{h}(b,0)=0 $ for $ b \in\mathbb{F}_{q}^{*}. $
It follows from Lemma \ref{lem6} that
\begin{equation} \label{eq-wt1}
S_{h}(b,1)=S_{h}(1,c^{-1})=\left\{\begin{array}{ll}
                                                                          0, & \textrm{if\ } \ \tr_{h}(c^{-1})\neq 1, \\
                                                                          \pm 2^{\frac{m+h}{2}}, & \textrm{if\ } \ \tr_{h}(c^{-1})= 1,
                                                                        \end{array}
                                                                        \right.
\end{equation}
where $c^{2^{h}+1}=b$ and $c\in \mathbb{F}_{q}^{*}. $ Together with equation \eqref{eq-weight}, we get
$$ |N(0,b)| \in \left\{2^{m-2},2^{m-2}-2^{\frac{m+h-4}{2}},2^{m-2}+2^{\frac{m+h-4}{2}}\right\}. $$
Hence,
 $$ \wt(c_{b})=2^{m-1}-|N(0,b)|\in \left\{2^{m-2},2^{m-2}\pm2^{\frac{m+h-4}{2}}\right\}.$$
Suppose
\begin{align*}
w_{1}=2^{m-2}-2^{\frac{m+h-4}{2}},\
 w_{2}=2^{m-2},\
 w_{3}=2^{m-2}+2^{\frac{m+h-4}{2}}.
\end{align*}
Note that $\gcd(2^{h}+1,2^{m}-1)=1$ if $m/h$ is odd. When $b$ ranges over $\mathbb{F}_{q}^{*}$, the element $c$ $(c^{2^{h}+1}=b)$ takes on each element of $\mathbb{F}_{q}^{*}$ exactly $1$ time. Hence, for  $b\in\mathbb{F}_{q}^{*}$ we obtain 
$$\left|\left\{c\in \mathbb{F}_{q}:\tr(c^{-1})\neq1,\ c^{2^{h}+1}=b\right\}\right|=2^{m}-2^{m-h}-1,$$
i.e., $A_{w_{2}}=2^{m}-2^{m-h}-1$.  The first two Pless Power Moments (\cite{HP03}, P.260)
yield  the following two equations:
\begin{eqnarray}\label{eq-a3}
\left\{\begin{array}{l}
         A_{w_1}+A_{w_2}+A_{w_3}=2^{m}-1, \\
         w_1A_{w_1}+w_2A_{w_2}+w_3A_{w_3}=n2^{m-1},
       \end{array}
       \right.
\end{eqnarray}
where $n=2^{m-1}-1. $ Solving the system of equations in  \eqref{eq-a3} gets Theorem \ref{thm1}.

The proof of Theorem \ref{thm2} is similar to that of Theorem \ref{thm1} and we omit the details.

In the sequel, we assume $ m/h\equiv0\pmod{2}$, $m=2e$ and $ g $
is a generator of $\mathbb{F}_{q}^{*}.$
In order to give the proof of Theorem \ref{thm3},  the following auxiliary  lemma is needed.
This lemma can be found in  equation (10) in \cite{DD14}.
\begin{lemma}\label{lem9}
Let $ T_{0}=|\{x\in\mathbb{F}_{q}:\tr(x^{2^{h}+1})=0\}|$ and $T_{1}=|\{x\in\mathbb{F}_{q}:\tr(x^{2^{h}+1})=1\}|. $
If $m/h $ is even, then $ T_{0}=2^{m-1}-(-1)^{\frac{e}{h}}2^{e+h-1}$ and $T_{1}=2^{m-1}+(-1)^{\frac{e}{h}}2^{e+h-1}.$
\end{lemma}
\vskip 2mm
\noindent{\textbf{Proof of Theorem \ref{thm3}.} }
If $b\in\mathbb{F}_{q}^{*}$ and $b\neq (g^{2^{h}+1})^{t} $ for any integer $t$, then
by Lemma \ref{lem7}, we have $$S_{h}(b,0)=(-1)^{\frac{e}{h}}2^{e},$$
and by Lemma \ref{lem8}, we get
 $$S_{h}(b,1)=(-1)^{\frac{e}{h}}2^{e}\chi_{1}(bx_{0}^{2^{h}+1}), $$
 where $b^{2^{h}}x_{0}^{2^{2h}}+bx_{0}=1.$

 If $ b= (g^{2^{h}+1})^{t} $
for some integer $t,$ it follows from Lemma \ref{lem7} that
$$S_{h}(b,0)=-(-1)^{\frac{e}{h}}2^{e+h}. $$
Assume $c=g^{t}$, then $b=c^{2^{h}+1}$ and by Lemma \ref{lem6} $S_{h}(b,1)=S_{h}(1,c^{-1}).$ For the above $c\in\mathbb{F}_{q}^{*}$,  let
\begin{equation}\label{fx}
f_c(x)=x^{2^{2h}}+x-(c^{-1})^{2^{h}}.
\end{equation}
  If $f_c(x)$ has no root in $\mathbb{F}_{q}^{*}, $ by Lemma \ref{lem8}, we obtain
$$S_{h}(b,1)=S_{h}(1,c^{-1})=0.$$
Note that $\tr_{h}(1)=0$, since $ m/h$ is even.
If $f_c(x) $ has a root $x_0$ in  $ \mathbb{F}_{q}^{*}, $ by Lemma \ref{lem8}, we get

$$S_{h}(b,1)=S_{h}(1,c^{-1})=-(-1)^{\frac{e}{h}}2^{e+h}\chi_{1}(x_{0}^{2^{h}+1}) .$$
Together with \eqref{eq-wt} and \eqref{eq-weight}, we know that  for $b\in \mathbb{F}_{q}^{*}$,
$$wt(\mathbf{c}_b)\in\left\{2^{m-2}+(-1)^{\frac{e}{h}}2^{e+h-1}, 2^{m-2}+(-1)^{\frac{e}{h}}2^{e+h-2}, 2^{m-2}, 2^{m-2}-(-1)^{\frac{e}{h}}2^{e-1} \right\}.$$

Define
$$
w_1=2^{m-2}+(-1)^{\frac{e}{h}}2^{e+h-1}, \ w_2=2^{m-2}+(-1)^{\frac{e}{h}}2^{e+h-2},\ w_3=2^{m-2}, \ w_4=2^{m-2}-(-1)^{\frac{e}{h}}2^{e-1}.
$$
The next step is to determine the number $A_{w_i}$ of codewords with weight $w_i$.
 If $f_c(x)=0$ (for some $c\in\mathbb{F}_{q})$  is solvable in $\mathbb{F}_{q}, $ by Lemma \ref{lem3.1},  there are $2^{2h}$ solutions of this equation  over $\mathbb{F}_{q}$. It can be easily checked that
 $$
 \left\{x_0\in\mathbb{F}_{q}: x_0^{2^{2h}}+x_0=(c^{-1})^{2^{h}}, c\in \mathbb{F}_{q} \right\}=\mathbb{F}_{q}.
 $$
 Hence we get
\begin{eqnarray*}
\left|\left\{c\in\mathbb{F}_{q}^{*}:  x^{2^{2h}}+x=(c^{-1})^{2^{h}} \textrm{\ is\ solvable\ in\ } \mathbb{F}_{q}\right\}\right|=2^{m-2h}-1,
\end{eqnarray*}
and
\begin{eqnarray*}
\left|\left\{c\in\mathbb{F}_{q}^{*}:  x^{2^{2h}}+x=(c^{-1})^{2^{h}} \textrm{\ has\ no\ root\ in\ } \mathbb{F}_{q}\right\}\right|=2^{m}-2^{m-2h}.
\end{eqnarray*}
Since $x^{2^{h}+1}$ is a $ (2^{h}+1)$-to-$1$ function on $ \mathbb{F}_{q}$, there are $\frac{2^{m}-2^{m-2h}}{2^{h}+1}$  $b$'s ($ b=c^{2^{h}+1}\in\mathbb{F}_{q}^{*}$) such that $S_{h}(b,1)=0$, i.e.,  $A_{w_{2}}=\frac{2^{m}-2^{m-2h}}{2^{h}+1}.$
It follows from Lemmas \ref{lem8} and \ref{lem9} that
 $$\left|\left\{c\in\mathbb{F}_{q}^{*}:S_{h}(1,c^{-1})=(-1)^{\frac{e}{h}}2^{e+h}\right\}\right|=\frac{2^{m-1}+(-1)^{\frac{e}{h}}2^{e+h-1}}{2^{2h}}.$$
Then we have
 $$
 \left|\left\{b\in\mathbb{F}_{q}^{*}:b=c^{2^{h}+1}\textrm{ and } S_{h}(b,1)=(-1)^{\frac{e}{h}}2^{e+h}\right\}\right|=\frac{2^{m-1}+(-1)^{\frac{e}{h}}2^{e+h-1}}{2^{2h}(2^{h}+1)}.
 $$
 Therefore,
$A_{w_{1}}=\frac{2^{m-2h-1}-1-(-1)^{\frac{e}{h}}2^{e-h-1}}{2^{h}+1}.$
By the Pless Power Moments (\cite{HP03}, p. 260) we obtain the following two equations:
\begin{eqnarray}\label{eq-a4}
\left\{\begin{array}{l}
         A_{w_1}+A_{w_2}+A_{w_3}+A_{w_4}=2^{m}-1, \\
         w_1A_{w_1}+w_2A_{w_2}+w_3A_{w_3}+w_4A_{w_4}=2^{m-1}(2^{m}-1).
       \end{array}
       \right.
\end{eqnarray}
The solutions of \eqref{eq-a4} yield the weight distribution of \autoref{tab3}.
The proof of Theorem \ref{thm3} is completed.

We omit the proof of Theorems \ref{thm4} and \ref{thm5}, since it is similar to that of Theorem \ref{thm3}.
\section{Concluding Remarks}
In this paper, we present a class of binary linear codes with no more than four weights.
A number of linear codes with  at most  five-weight codes were discussed in \cite{CW84,Ding09,DD14,DD15,DGZ13,ZD14,ZLFH15}.

It should be remarked that  he parameters of the binary linear codes in Theorem \ref{thm1} are
the same as those in Theorem 1 in \cite{DD14}. It is open whether the two class of codes are
equivalent. The readers are  invited to attack this problem.

 Denote the minimum and maximum nonzero weight of a linear code $\C$ over $\mathbb{F}_{p}$ by $w_{\min}$ and $w_{\max},$ respectively. By the results in \cite{YD06}, if the code $\C$ satisfies the following inequality
 $$
 \frac{w_{\min}}{w_{\max}}> \frac{p-1}{p},
 $$
 then $\C$ can be employed to construct secret sharing schemes with interesting properties.

Let $m > h+2. $ Then for the codes in Theorems \ref{thm1} and \ref{thm2}, we have
$$
\frac{w_{\min}}{w_{\max}}=\frac{2^{m-2}-2^{\frac{m+h-4}{2}}}{2^{m-2}+2^{\frac{m+h-4}{2}}} >\frac{1}{2}.
$$
If $(m,h)\neq(4,1)$ or $(6,1), $ then for the codes in Theorems \ref{thm3} and \ref{thm4}, it can be easily checked that
$$
\frac{w_{\min}}{w_{\max}}> \frac{1}{2}.
$$
This conclusion is true also for Theorem \ref{thm5} and Corollary \ref{corollary6}.

Hence, the binary linear codes presented in this paper are suitable for constructing secret sharing schemes in many cases.


\begin{thebibliography}{99}
\bibitem{BK91}
 I. F. Blake and K. Kith, On the complete weight enumerator of Reed-Solomon codes, \emph{SIAM J. Disc. Math.}, vol. 4, no. 2, pp. 164--171, 1991.
 \bibitem{C79}
 L. Carlitz, Exolicite evaluation of certain exponential sums, \emph{Math. Scand,} vol. 44, pp. 5--16, 1979.
 \bibitem{C80}
L. Carlitz, Evaluation of some exponential sums over a finite field, \emph{Math. Nachr,} vol. 96, pp. 319--339, 1980.
 \bibitem{C98}
 R. S. Coulter, Explicit evaluation of some Weil sums, \emph{Acta Arith.,} vol. 83, pp. 241--251, 1998.
\bibitem{C99} R. S. Coulter, ¡°On the evaluation of a class of Weil sums in characteristic 2,¡±  \emph{New Zealand J. of Math.,} vol. 28, pp. 171--184, 1999.
\bibitem{CKNC12}
 S. -T. Choi, J. -Y. Kim, J. -S. No, and H. Chung, Weight distribution of
some cyclic codes, in \emph{Proc. Int. Symp. Inf. Theory,}  pp. 2911--2913, 2012.

\bibitem{CW84}
 B. Courteau, J. Wolfmann,  On triple--sum--sets and two or three weights codes,  \emph{Discret. Math.,} vol. 50,  pp. 179--191,  1984.
 \bibitem{Ding09}
 C. Ding, A class of three-weight and four-weight codes, in: C. Xing, et al. (Eds.), Proc. of
the Second International Workshop on Coding Theory and Cryptography, in: Lecture Notes
in Computer Science, vol. 5557, Springer Verlag, pp. 34--42, 2009.
\bibitem{Ding15}
 C. Ding, Linear codes from some 2-designs, \emph{IEEE Trans. Inf. Theory,} vol.  61, no. 6, pp. 3265-3275, 2015.
\bibitem{DD14}
 K. Ding,  C. Ding, Bianry linear codes with three weights, \emph{IEEE Commun. Lett.},
vol. 18, no. 11, pp. 1879--1882,  2014.
\bibitem{DD15}
K. Ding,  C. Ding, A class of two-weight and three-weight codes and their applications in secret sharing, \emph{IEEE Trans.  Inf. Theory}, vol. 61, no. 11, pp. 5835--5842, 2015.
 \bibitem{DGZ13}
C. Ding, Y. Gao, Z. Zhou,
Five Families of Three-Weight Ternary Cyclic Codes and Their Duals, \emph{IEEE Trans.  Inf. Theory}, vol. 59, no. 12, pp. 7940--7946, 2013.
 \bibitem{DLMZ11}
  C. Ding, Y. Liu, C. Ma, L. Zeng, The weight distributions of the duals of cyclic codes with two zeros, \emph{IEEE Trans. Inf. Theory,} vol. 57, no. 12, pp. 8000--8006, 2011.
 \bibitem{DLN08}
C. Ding, J. Luo, H. Niederreiter, Two-weight codes punctured from irreducible cyclic codes,
in: Y. Li, et al. (Eds.), Proceedings of the First Worshop on Coding and Cryptography,
World Scientific, Singapore, pp. 119--124, 2008.
\bibitem{DLLZ15}
C. Ding, C. Li, N. Li, and Z. Zhou, Three-weight cyclic codes and their
weight distributions, \emph{Disctret. Math.,} vol. 339, no. 2, pp. 415--427, 2016.
\bibitem{DY13}
C. Ding, J. Yang, Hamming weights in irreducible cyclic codes, \emph{Discret. Math.,} vol. 313, no. 4,  pp. 434--446,  2013.

\bibitem{DH07}
C. Ding,  H. Niederreiter,
Cyclotomic linear codes of order $3$, emph{IEEE Trans.  Inf. Theory}, vol. 53, no. 6, pp. 2274--2277, 2007.

\bibitem{F12}
 T. Feng, On cyclic codes of length $2^{2^r}-1$ with two zeros whose dual codes have three weights. Des. Codes Cryptogr. vol. 62, pp. 253--258, 2012.
 \bibitem{FL08}
  K. Feng, J. Luo, Weight distribution of some reducible cyclic codes. \emph{Finite Fields Appl.,} vol. 14, no. 2, pp. 390--409, 2008.
  \bibitem{FLX12}
  T. Feng, K. Leung, Q. Xiang, Binary cyclic codes with two primitive nonzeros, \emph{Sci. China Math.,} vol. 56, no. 7, pp. 1403--1412, 2012.
   \bibitem{H07}
 X. Hou, Explicit evaluation of certain exponential sums of binary quadratic functions, \emph{Finite Fields Appl.,} 13, 843--868, 2007.
\bibitem{HP03}
 W. C. Huffman and V. Pless, {\em Fundamentals of error-correcting codes, }
Cambridge: Cambridge University Press, 2003.
\bibitem{K89}
 K. Kith, Complete weight enumeration of Reed-Solomon codes, Master¡¯s thesis, Department of Electrical and Computing Engineering, University of Waterloo, Waterloo, Ontario, Canada, 1989.
\bibitem{LF08}
 J. Luo, K. Feng, On the weight distribution of two classes of cyclic codes. \emph{IEEE Trans. Inf. Theory,} vol.54, no. 12, pp. 5332--5344, 2008.
\bibitem{LYL14}
 C. Li, Q. Yue, and F. Li, Hamming weights of the duals of cyclic codes
with two zeros, \emph{IEEE Trans. Inf. Theory,} vol. 60, no. 7, pp. 3895--3902, Jul. 2014.
\bibitem{LN97}
 R. Lidl, H. Niederreiter, { Finite fields}. Cambridge University Press, New York (1997)
\bibitem{WDX15}
Q. Wang, K. Ding, R. Xue, Binary linear codes with two weights, \emph{IEEE Commun. Lett.}, vol. 19, no. 7, Jul. 2015.
\bibitem{WDLX15}
Q. Wang, K. Ding, D. Lin, R. Xue, A kind of three-weight linear codes, DOI 10.1007/s12095-015-0180-3
 \bibitem{YCD06}
J. Yuan, C. Carlet, C. Ding, The weight distribution of a class of linear codes from perfect nonlinear functions, \emph{IEEE Trans. Inf. Theory,} vol. 52, no. 2, 712--717, 2006.
\bibitem{YD06}
J. Yuan, C. Ding, Secret sharing schemes from three classes of linear codes,
\emph{ IEEE Trans. Inf. Theory,} vol. 52, no. 1, pp. 206--212, 2006.
\bibitem{ZD14}
Z. Zhou, C. Ding,  A class of three--weight cyclic codes, \emph{Finite Fields and Their Appl.,} vol. 25, pp. 79--93, 2014.
\bibitem{ZLFH15}
Z. Zhou, N. Li, C. Fan, T. Helleseth,  Linear codes with two or three weights from quadratic bent functions, DOI 10.1007/s10623-015-0144-9
\end{thebibliography}
\end{document}